\documentclass[12pt]{article}
\usepackage{axodraw,bbold}

\catcode`@=12
\begin{document}
\thispagestyle{empty}
\begin{flushright} 
UCRHEP-T426\\ 
December 2006\
\end{flushright}
\vspace{0.5in}
\begin{center}
{\LARGE	\bf Lepton Family Symmetry and Possible\\
Application to the Koide Mass Formula\\}
\vspace{1.5in}
{\bf Ernest Ma\\}
\vspace{0.2in}
{\sl Physics and Astronomy Department\\ University of California, Riverside\\ 
Riverside, California 92521, USA \\}
\vspace{1.5in}
\end{center}

\begin{abstract}\
A finite group generated by four $Z_3$ transformations is applied to 
lepton families in a supersymmetric model, resulting in the charged-lepton 
masses $m_i$ being proportional to $v_i^2$, where $v_i$ are three vacuum 
expectation values.  This may be relevant in obtaining the Koide formula 
$m_e+m_\mu+m_\tau = (2/3)(\sqrt{m_e}+\sqrt{m_\mu}+\sqrt{m_\tau})^2$.
\end{abstract}

\newpage
\baselineskip 24pt
The Koide mass formula \cite{k82,k83,k90,k05}
\begin{equation}
m_e + m_\mu + m_\tau = {2 \over 3} \left( \sqrt{m_e} + \sqrt{m_\mu} + 
\sqrt{m_\tau} \right)^2
\end{equation}
is remarkably accurate, and cries out for a possible theoretical explanation.
Koide himself has proposed \cite{k06-1} to understand it as
\begin{equation}
v_1^2 + v_2^2 + v_3^2 = {2 \over 3} (v_1 + v_2 + v_3)^2.
\end{equation}
In that case, the condition $m_i \propto v_i^2$ is required, and must be valid 
to great precision.  To facilitate such a relationship, the following 
non-Abelian discrete family symmetry is proposed.

Consider the basis $(a_1,a_2,a_3)$ and four separate $Z_3$ transformations:
\begin{equation}
a_1 \to a_2 \to a_3 \to a_1, ~~~ a_i \to \omega a_i \to \omega^2 a_i \to a_i, 
~(i=1,2,3),
\end{equation}
where $\omega = \exp(2 \pi i/3) = -1/2 + i \sqrt 3/2$.  The resulting group, 
call it $\Sigma(81)$, has 81 elements, divided into 17 equivalence classes. 
It has 9 one-dimensional irreducible representations 
${\bf 1_i} (i=1,...,9$) and 8 three-dimensional ones ${\bf 3_A}, 
{\bf \bar{3}_A}$, ${\bf 3_B}, {\bf \bar{3}_B}$, ${\bf 3_C}, {\bf \bar{3}_C}$, 
${\bf 3_D}, {\bf \bar{3}_D}$.  Its character table is given in the Appendix, 
together with the 81 matrices of the defining representation ${\bf 3_A}$.

The key property of $\Sigma(81)$ which allows this to work is that given just 
the representations ${\bf 3_A}$ and ${\bf \bar{3}_A}$, there are only 
three invariants:
\begin{equation}
a_1 a_1 a_1 + a_2 a_2 a_2 + a_3 a_3 a_3, ~~~ a_1 \bar a_1 + a_2 \bar a_2 + 
a_3 \bar a_3, ~~~ \bar a_1 \bar a_1 \bar a_1 + \bar a_2 \bar a_2 \bar a_2 + 
\bar a_3 \bar a_3 \bar a_3.
\end{equation}
Consider now the supersymmetric standard model, with lepton superfields
\begin{equation}
L_i = (\nu_i,l_i), ~~ l^c_i \sim {\bf 3_A}.
\end{equation}
Add the following Higgs superfields:
\begin{eqnarray}
&& \eta_i = (\eta^0_i,\eta^-_i), ~~ \xi_i = (\xi^+_i,\xi^0_i) \sim {\bf 3_A}, 
\\
&& \zeta_i = (\zeta^0_i,\zeta^-_i), ~~ \psi_i = (\psi^+_i,\psi^0_i) \sim 
{\bf \bar{3}_A}, \\
&& \sigma_i = \sigma^0_i \sim {\bf \bar{3}_A}, ~~~ \phi = (\phi^0,\phi^-) 
\sim {\bf 1}.
\end{eqnarray}
The relevant superpotential is then given by
\begin{equation}
W = f L_i l^c_i \eta_i + M_1 \eta_i \psi_i + h_1 \zeta_i \psi_i \sigma_i + 
M_2 \zeta_i \xi_i + h_2 \xi_i \sigma_i \phi,
\end{equation}
resulting in the scalar potential
\begin{equation}
V = |M_1 \psi_i + f L_i l^c_i|^2 + |M_2 \xi_i + h_1 \psi_i \sigma_i|^2 + 
|M_1 \eta_i + h_1 \zeta_i \sigma_i|^2 + |M_2 \zeta_i + h_2 \sigma_i \phi|^2.
\end{equation}
Assuming  $\langle \phi^0 \rangle \neq 0$ and $\langle \sigma_i \rangle 
\neq 0$ (to be discussed later), the supersymmetric minimum $V=0$  
is obtained for
\begin{equation}
\langle \psi_i^0 \rangle = \langle \xi_i^0 \rangle = 0, ~~~ 
\langle \eta^0_i \rangle = {- h_1 \langle \zeta_i^0 \rangle \langle \sigma_i 
\rangle \over M_1}, ~~~ 
\langle \zeta^0_i \rangle = {- h_2 \langle \sigma_i \rangle \langle \phi^0 
\rangle \over M_2},
\end{equation}
resulting thus in the charged-lepton mass relationship
\begin{equation}
m_i = f \langle \eta^0_i \rangle = {f h_1 h_2 \langle \sigma_i \rangle^2 
\langle \phi^0 \rangle \over M_1 M_2},
\end{equation}
exactly as advertised.  

To obtain Eq.~(2), let \cite{k06-1}
\begin{equation}
\pmatrix{\chi_0 \cr \chi_1 \cr \chi_2} = {1 \over \sqrt{3}} \pmatrix{1 & 1 & 
1 \cr 1 & \omega^2 & \omega \cr 1 & \omega & 
\omega^2} \pmatrix{\sigma_1 \cr \sigma_2 \cr \sigma_3},
\end{equation}
so that
\begin{eqnarray}
\sigma_1^2 + \sigma_2^2 + \sigma_3^2 &=& \chi_0^2 + 2 \chi_1 \chi_2, \\ 
\sigma_1^3 + \sigma_2^3 + \sigma_3^3 &=& {1 \over \sqrt{3}} (\chi_0^3 + 
\chi_1^3 + \chi_2^3 + 6 \chi_0 \chi_1 \chi_2). 
\end{eqnarray}
Consider the superpotential
\begin{equation}
W = {1 \over 2} m_0 \chi_0^2 + {1 \over 2} m_1 \chi_1^2 + {1 \over 2} m_2 
\chi_2^2 + m_3 \chi_1 \chi_2 + {1 \over 3} \lambda (\chi_0^3 + \chi_1^3 + 
\chi_2^3 + 6 \chi_0 \chi_1 \chi_2),
\end{equation}
which breaks $\Sigma(81)$ only softly with its bilinear terms. 
The resulting scalar potential is given by
\begin{eqnarray}
V &=& |m_0 \chi_0 + \lambda (\chi_0^2 + 2 \chi_1 \chi_2)|^2 + 
|m_1 \chi_1 + m_3 \chi_2 + \lambda (\chi_1^2 + 2 \chi_0 \chi_2)|^2 \nonumber \\
&+& |m_2 \chi_2 + m_3 \chi_1 + \lambda (\chi_2^2 + 2 \chi_0 \chi_1)|^2.
\end{eqnarray}

The aim now is to find a supersymmetric minimum, i.e. $V=0$, for which
\begin{equation}
\langle \chi_0 \rangle^2 = 2 \langle \chi_1 \rangle \langle \chi_2 
\rangle = {1 \over 2} (\langle \chi_0 \rangle^2 + 2 \langle \chi_1 
\rangle \langle \chi_2 \rangle),
\end{equation}
which is equivalent to \cite{k06-1}
\begin{equation}
{1 \over 3} (\langle \sigma_1 \rangle + \langle \sigma_2 \rangle + \langle 
\sigma_3 \rangle)^2 =  
{1 \over 2} (\langle \sigma_1 \rangle^2 + \langle \sigma_2 
\rangle^2 + \langle \sigma_3 \rangle^2),
\end{equation}
i.e. Eq.~(2), and then use Eq.~(12) to obtain the Koide formula of Eq.~(1).
This turns out to be extremely simple, i.e.
\begin{equation}
m_3 = m_0 = \sqrt{8m_1m_2},
\end{equation}
for which
\begin{equation}
\langle \chi_0 \rangle = {-m_0 \over 2 \lambda}, ~~~ 
\langle \chi_1 \rangle = {-m_1 \over \lambda}, ~~~ 
\langle \chi_2 \rangle = {-m_2 \over \lambda}. 
\end{equation}

The validity of Eq.~(20) is of course not understood.  It may be due 
to some underlying symmetry or dynamics, but whatever the origin, once it is 
established, Eq.~(2) is protected by the unbroken supersymmetry of the theory. 
Assuming that $\chi_i$ acquire vaccum expectation values not far above 
the supersymmetry breaking scale, the effective Yukawa couplings of 
the leptons at the electroweak scale will not deviate very much from 
their values which predict Eq.~(1).

As for the neutrino sector \cite{k06-2}, it is clear that having only 
superfields which transform as ${\bf 3_A}$ and ${\bf \bar{3}_A}$ will 
not work, because that necessarily yields a diagonal mass matrix. 
To obtain a realistic neutrino mass matrix, consider for example scalar 
superfields $(\Delta_i^{++},\Delta_i^+,\Delta_i^0)$ which transform as 
${\bf 3_A}$ and ${\bf 3_B}$ under $\Sigma(81)$.  Using the multiplication 
rule
\begin{equation}
{\bf 3_A} \times {\bf 3_A} = {\bf \bar{3}_A} + {\bf \bar{3}_B} + 
{\bf \bar{3}_B}, 
\end{equation}
it is possible to have ${\cal M}_\nu$ of the most general form, i.e.
\begin{equation}
{\cal M}_\nu = \pmatrix{a & f & e \cr f & b & d \cr e & d & c},
\end{equation}
where $(a,b,c)$ come from ${\bf 3_A}$ and $(d,e,f)$ from ${\bf 3_B}$. 
This makes no prediction, but if the vacuum structure can be restricted, 
say $b=c$ and $e=f$, as for example in a model \cite{m06-1} based on $S_4$, 
then the approximate tribimaximal mixing of neutrinos may be obtained.

\noindent {\bf Acknowledgements}

I thank Yoshio Koide for correspondence.  This work was supported in part 
by the U.~S.~Department of Energy under Grant No. DE-FG03-94ER40837.\\[5pt]

\newpage
\noindent {\bf Appendix} 

The proposed finite group $\Sigma(81)$ contains $\Delta(27)$ 
\cite{bgg84,mvkr06}.  The latter's character table and defining 
representation have been given elsewhere \cite{m06-2}.  Here the 
$17 \times 17$ character table of $\Sigma(81)$ and the 81 matrices 
of its defining representation ${\bf 3_A}$ are also given.

\begin{table}[htb]
\caption{Character table of $\Sigma(81)$ for the 9 one-dimensional 
representations; $n$ is the number of elements and $h$ is the order of each 
element.}

\begin{center}
\begin{tabular}{|c|c|c|c|c|c|c|c|c|c|c|c|}
\hline 
Class & $n$ & $h$ & ${\bf 1_1}$ & ${\bf 1_2}$ & ${\bf 1_3}$ & ${\bf 1_4}$ 
& ${\bf 1_5}$ & ${\bf 1_6}$ & ${\bf 1_7}$ & ${\bf 1_8}$ & ${\bf 1_9}$ \\ 
\hline
$C_1$ & 1 & 1 & 1 & 1 & 1 & 1 & 1 & 1 & 1 & 1 & 1 \\ 
$C_2$ & 1 & 3 & 1 & 1 & 1 & 1 & 1 & 1 & 1 & 1 & 1 \\ 
$C_3$ & 1 & 3 & 1 & 1 & 1 & 1 & 1 & 1 & 1 & 1 & 1 \\ 
$C_4$ & 3 & 3 & 1 & 1 & 1 & 1 & 1 & 1 & 1 & 1 & 1 \\ 
$C_5$ & 3 & 3 & 1 & 1 & 1 & 1 & 1 & 1 & 1 & 1 & 1 \\ 
$C_6$ & 9 & 3 & 1 & $\omega$ & $\omega^2$ & 1 & $\omega^2$ & $\omega$ & 1 & 
$\omega$ & $\omega^2$ \\ 
$C_7$ & 9 & 3 & 1 & $\omega^2$ & $\omega$ & 1 & $\omega$ & $\omega^2$ & 1 & 
$\omega^2$ & $\omega$ \\ 
$C_8$ & 3 & 3 & 1 & 1 & 1 & $\omega^2$ & $\omega^2$ & $\omega^2$ & $\omega$ & 
$\omega$ & $\omega$ \\ 
$C_9$ & 3 & 3 & 1 & 1 & 1 & $\omega^2$ & $\omega^2$ & $\omega^2$ & $\omega$ & 
$\omega$ & $\omega$ \\ 
$C_{10}$ & 3 & 3 & 1 & 1 & 1 & $\omega^2$ & $\omega^2$ & $\omega^2$ & 
$\omega$ & $\omega$ & $\omega$ \\ 
$C_{11}$ & 3 & 3 & 1 & 1 & 1 & $\omega$ & $\omega$ & $\omega$ & $\omega^2$ & 
$\omega^2$ & $\omega^2$ \\ 
$C_{12}$ & 3 & 3 & 1 & 1 & 1 & $\omega$ & $\omega$ & $\omega$ & $\omega^2$ & 
$\omega^2$ & $\omega^2$ \\ 
$C_{13}$ & 3 & 3 & 1 & 1 & 1 & $\omega$ & $\omega$ & $\omega$ & $\omega^2$ & 
$\omega^2$ & $\omega^2$ \\ 
$C_{14}$ & 9 & 9 & 1 & $\omega$ & $\omega^2$ & $\omega^2$ & $\omega$ & 1 & 
$\omega$ & $\omega^2$ & 1 \\ 
$C_{15}$ & 9 & 9 & 1 & $\omega^2$ & $\omega$ & $\omega^2$ & 1 & $\omega$ & 
$\omega$ & 1 & $\omega^2$ \\ 
$C_{16}$ & 9 & 9 & 1 & $\omega^2$ & $\omega$ & $\omega$ & $\omega^2$ & 1 
& $\omega^2$ & $\omega$ & 1 \\ 
$C_{17}$ & 9 & 9 & 1 & $\omega$ & $\omega^2$ & $\omega$ & 1 & $\omega^2$ 
& $\omega^2$ & 1 & $\omega$ \\ 
\hline
\end{tabular}
\end{center}
\end{table}

\newpage
\begin{table}[htb]
\caption{Character table of $\Sigma(81)$ for the 8 three-dimensional 
representations.}
\begin{center}
\begin{tabular}{|c|c|c|c|c|c|c|c|c|c|c|}
\hline 
Class & $n$ & $h$ & ${\bf 3_A}$ & ${\bf \bar{3}_A}$ & ${\bf 3_B}$ & 
${\bf \bar{3}_B}$ & ${\bf 3_C}$ & ${\bf \bar{3}_C}$ & ${\bf 3_D}$ & 
${\bf \bar{3}_D}$ \\ 
\hline
$C_1$ & 1 & 1 & 3 & 3 & 3 & 3 & 3 & 3 & 3 & 3 \\ 
$C_2$ & 1 & 3 & 3$\omega$ & 3$\omega^2$ & 3$\omega$ & 3$\omega^2$ & 3$\omega$ 
& 3$\omega^2$ & 3 & 3 \\ 
$C_3$ & 1 & 3 & 3$\omega^2$ & 3$\omega$ & 3$\omega^2$ & 3$\omega$ & 
3$\omega^2$ & 3$\omega$ & 3 & 3 \\ 
$C_4$ & 3 & 3 & 0 & 0 & 0 & 0 & 0 & 0 & 3$\omega$ & 3$\omega^2$ \\ 
$C_5$ & 3 & 3 & 0 & 0 & 0 & 0 & 0 & 0 & 3$\omega^2$ & 3$\omega$ \\ 
$C_6$ & 9 & 3 & 0 & 0 & 0 & 0 & 0 & 0 & 0 & 0 \\ 
$C_7$ & 9 & 3 & 0 & 0 & 0 & 0 & 0 & 0 & 0 & 0 \\
$C_8$ & 3 & 3 & $-i\sqrt{3}$ & $i\sqrt{3}$ & $-i\sqrt{3}\omega^2$ & 
$i\sqrt{3}\omega$ & $-i\sqrt{3}\omega$ & $i\sqrt{3}\omega^2$ & 0 & 0 \\ 
$C_9$ & 3 & 3 & $-i\sqrt{3}\omega$ & $i\sqrt{3}\omega^2$ & $-i\sqrt{3}$ & 
$i\sqrt{3}$ & $-i\sqrt{3}\omega^2$ & $i\sqrt{3}\omega$ & 0 & 0 \\ 
$C_{10}$ & 3 & 3 & $-i\sqrt{3}\omega^2$ & $i\sqrt{3}\omega$ & 
$-i\sqrt{3}\omega$ & $i\sqrt{3}\omega^2$ & $-i\sqrt{3}$ & $i\sqrt{3}$ & 
0 & 0 \\ 
$C_{11}$ & 3 & 3 & $i\sqrt{3}$ & $-i\sqrt{3}$ & $i\sqrt{3}\omega$ & 
$-i\sqrt{3}\omega^2$ & $i\sqrt{3}\omega^2$ & $-i\sqrt{3}\omega$ & 0 & 0 \\ 
$C_{12}$ & 3 & 3 & $i\sqrt{3}\omega$ & $-i\sqrt{3}\omega^2$ & 
$i\sqrt{3}\omega^2$ & $-i\sqrt{3}\omega$ & $i\sqrt{3}$ & $-i\sqrt{3}$ & 
0 & 0 \\ 
$C_{13}$ & 3 & 3 & $i\sqrt{3}\omega^2$ & $-i\sqrt{3}\omega$ & $i\sqrt{3}$ & 
$-i\sqrt{3}$ & $i\sqrt{3}\omega$ & $-i\sqrt{3}\omega^2$ & 0 & 0 \\ 
$C_{14}$ & 9 & 9 & 0 & 0 & 0 & 0 & 0 & 0 & 0 & 0 \\ 
$C_{15}$ & 9 & 9 & 0 & 0 & 0 & 0 & 0 & 0 & 0 & 0 \\ 
$C_{16}$ & 9 & 9 & 0 & 0 & 0 & 0 & 0 & 0 & 0 & 0 \\ 
$C_{17}$ & 9 & 9 & 0 & 0 & 0 & 0 & 0 & 0 & 0 & 0 \\ 
\hline
\end{tabular}
\end{center}
\end{table}

\newpage
The matrices of the ${\bf 3_A}$ representation 
of $\Sigma(81)$ are given by
\begin{eqnarray}
C_1 &:& \pmatrix{1 & 0 & 0 \cr 0 & 1 & 0 \cr 0 & 0 & 1}, ~~~
C_2 ~:~ \pmatrix{\omega & 0 & 0 \cr 0 & \omega & 0 \cr 0 & 0 & \omega}, ~~~
C_3 ~:~ \pmatrix{\omega^2 & 0 & 0 \cr 0 & \omega^2 & 0 \cr 0 & 0 & \omega^2}, 
\\ 
C_4 &:& \pmatrix{1 & 0 & 0 \cr 0 & \omega & 0 \cr 0 & 0 & \omega^2}, 
\pmatrix{\omega^2 & 0 & 0 \cr 0 & 1 & 0 \cr 0 & 0 & \omega}, 
\pmatrix{\omega & 0 & 0 \cr 0 & \omega^2 & 0 \cr 0 & 0 & 1}, \\
C_5 &:& \pmatrix{1 & 0 & 0 \cr 0 & \omega^2 & 0 \cr 0 & 0 & \omega},
\pmatrix{\omega & 0 & 0 \cr 0 & 1 & 0 \cr 0 & 0 & \omega^2}, 
\pmatrix{\omega^2 & 0 & 0 \cr 0 & \omega & 0 \cr 0 & 0 & 1}, \\
C_6 &:& \pmatrix{0 & 1 & 0 \cr 0 & 0 & 1 \cr 1 & 0 & 0}, 
\pmatrix{0 & \omega & 0 \cr 0 & 0 & \omega \cr \omega & 0 & 0}, 
\pmatrix{0 & \omega^2 & 0 \cr 0 & 0 & \omega^2 \cr \omega^2 & 0 & 0}, 
\pmatrix{0 & 1 & 0 \cr 0 & 0 & \omega \cr \omega^2 & 0 & 0},  
\pmatrix{0 & \omega^2 & 0 \cr 0 & 0 & 1 \cr \omega & 0 & 0}, \nonumber \\
&:& \pmatrix{0 & \omega & 0 \cr 0 & 0 & \omega^2 \cr 1 & 0 & 0}, 
\pmatrix{0 & 1 & 0 \cr 0 & 0 & \omega^2 \cr \omega & 0 & 0}, 
\pmatrix{0 & \omega & 0 \cr 0 & 0 & 1 \cr \omega^2 & 0 & 0}, 
\pmatrix{0 & \omega^2 & 0 \cr 0 & 0 & \omega \cr 1 & 0 & 0}, \\ 
C_7 &:& \pmatrix{0 & 0 & 1 \cr 1 & 0 & 0 \cr 0 & 1 & 0}, 
\pmatrix{0 & 0 & \omega \cr \omega & 0 & 0 \cr 0 & \omega & 0}, 
\pmatrix{0 & 0 & \omega^2 \cr \omega^2 & 0 & 0 \cr 0 & \omega^2 & 0}, 
\pmatrix{0 & 0 & 1 \cr \omega & 0 & 0 \cr 0 & \omega^2 & 0}, 
\pmatrix{0 & 0 & \omega^2 \cr 1 & 0 & 0 \cr 0 & \omega & 0}, \nonumber \\ 
&:& \pmatrix{0 & 0 & \omega \cr \omega^2 & 0 & 0 \cr 0 & 1 & 0}, 
\pmatrix{0 & 0 & 1 \cr \omega^2 & 0 & 0 \cr 0 & \omega & 0}, 
\pmatrix{0 & 0 & \omega \cr 1 & 0 & 0 \cr 0 & \omega^2 & 0}, 
\pmatrix{0 & 0 & \omega^2 \cr \omega & 0 & 0 \cr 0 & 1 & 0}, \\
C_8 &:& \pmatrix{\omega^2 & 0 & 0 \cr 0 & \omega^2 & 0 \cr 0 & 0 & 1},  
\pmatrix{\omega^2 & 0 & 0 \cr 0 & 1 & 0 \cr 0 & 0 & \omega^2}, 
\pmatrix{1 & 0 & 0 \cr 0 & \omega^2 & 0 \cr 0 & 0 & \omega^2}, \\
C_9 &:& \pmatrix{\omega & 0 & 0 \cr 0 & 1 & 0 \cr 0 & 0 & 1},  
\pmatrix{1 & 0 & 0 \cr 0 & \omega & 0 \cr 0 & 0 & 1}, 
\pmatrix{1 & 0 & 0 \cr 0 & 1 & 0 \cr 0 & 0 & \omega}, \\
C_{10} &:& \pmatrix{\omega & 0 & 0 \cr 0 & \omega & 0 \cr 0 & 0 & \omega^2},  
\pmatrix{\omega & 0 & 0 \cr 0 & \omega^2 & 0 \cr 0 & 0 & \omega}, 
\pmatrix{\omega^2 & 0 & 0 \cr 0 & \omega & 0 \cr 0 & 0 & \omega}, \\
C_{11} &:& \pmatrix{\omega & 0 & 0 \cr 0 & \omega & 0 \cr 0 & 0 & 1},  
\pmatrix{\omega & 0 & 0 \cr 0 & 1 & 0 \cr 0 & 0 & \omega}, 
\pmatrix{1 & 0 & 0 \cr 0 & \omega & 0 \cr 0 & 0 & \omega}, \\
C_{12} &:& \pmatrix{\omega^2 & 0 & 0 \cr 0 & \omega^2 & 0 \cr 0 & 0 & 
\omega},  \pmatrix{\omega^2 & 0 & 0 \cr 0 & \omega & 0 \cr 0 & 0 & \omega^2}, 
\pmatrix{\omega & 0 & 0 \cr 0 & \omega^2 & 0 \cr 0 & 0 & \omega^2}, \\
C_{13} &:& \pmatrix{\omega^2 & 0 & 0 \cr 0 & 1 & 0 \cr 0 & 0 & 1},  
\pmatrix{1 & 0 & 0 \cr 0 & \omega^2 & 0 \cr 0 & 0 & 1}, 
\pmatrix{1 & 0 & 0 \cr 0 & 1 & 0 \cr 0 & 0 & \omega^2}, \\
C_{14} &:& \pmatrix{0 & \omega & 0 \cr 0 & 0 & 1 \cr 1 & 0 & 0}, 
\pmatrix{0 & 1 & 0 \cr 0 & 0 & \omega \cr 1 & 0 & 0}, 
\pmatrix{0 & 1 & 0 \cr 0 & 0 & 1 \cr \omega & 0 & 0}, 
\pmatrix{0 & \omega^2 & 0 \cr 0 & 0 & \omega^2 \cr 1 & 0 & 0},  
\pmatrix{0 & \omega^2 & 0 \cr 0 & 0 & 1 \cr \omega^2 & 0 & 0}, \nonumber \\
&:& \pmatrix{0 & 1 & 0 \cr 0 & 0 & \omega^2 \cr \omega^2 & 0 & 0},  
\pmatrix{0 & \omega & 0 \cr 0 & 0 & \omega \cr \omega^2 & 0 & 0}, 
\pmatrix{0 & \omega & 0 \cr 0 & 0 & \omega^2 \cr \omega & 0 & 0}, 
\pmatrix{0 & \omega^2 & 0 \cr 0 & 0 & \omega \cr \omega & 0 & 0}, \\ 
C_{15} &:& \pmatrix{0 & \omega^2 & 0 \cr 0 & 0 & 1 \cr 1 & 0 & 0}, 
\pmatrix{0 & 1 & 0 \cr 0 & 0 & \omega^2 \cr 1 & 0 & 0}, 
\pmatrix{0 & 1 & 0 \cr 0 & 0 & 1 \cr \omega^2 & 0 & 0}, 
\pmatrix{0 & \omega & 0 \cr 0 & 0 & \omega \cr 1 & 0 & 0},   
\pmatrix{0 & \omega & 0 \cr 0 & 0 & 1 \cr \omega & 0 & 0}, \nonumber \\
&:& \pmatrix{0 & 1 & 0 \cr 0 & 0 & \omega \cr \omega & 0 & 0}, 
\pmatrix{0 & \omega^2 & 0 \cr 0 & 0 & \omega^2 \cr \omega & 0 & 0}, 
\pmatrix{0 & \omega^2 & 0 \cr 0 & 0 & \omega \cr \omega^2 & 0 & 0}, 
\pmatrix{0 & \omega & 0 \cr 0 & 0 & \omega^2 \cr \omega^2 & 0 & 0}, \\ 
C_{16} &:& \pmatrix{0 & 0 & \omega \cr 1 & 0 & 0 \cr 0 & 1 & 0}, 
\pmatrix{0 & 0 & 1 \cr \omega & 0 & 0 \cr 0 & 1 & 0}, 
\pmatrix{0 & 0 & 1 \cr 1 & 0 & 0 \cr 0 & \omega & 0}, 
\pmatrix{0 & 0 & \omega^2 \cr \omega^2 & 0 & 0 \cr 0 & 1 & 0}, 
\pmatrix{0 & 0 & \omega^2 \cr 1 & 0 & 0 \cr 0 & \omega^2 & 0}, \nonumber \\
&:& \pmatrix{0 & 0 & 1 \cr \omega^2 & 0 & 0 \cr 0 & \omega^2 & 0},  
\pmatrix{0 & 0 & \omega \cr \omega & 0 & 0 \cr 0 & \omega^2 & 0}, 
\pmatrix{0 & 0 & \omega \cr \omega^2 & 0 & 0 \cr 0 & \omega & 0}, 
\pmatrix{0 & 0 & \omega^2 \cr \omega & 0 & 0 \cr 0 & \omega & 0}, \\
C_{17} &:& \pmatrix{0 & 0 & \omega^2 \cr 1 & 0 & 0 \cr 0 & 1 & 0}, 
\pmatrix{0 & 0 & 1 \cr \omega^2 & 0 & 0 \cr 0 & 1 & 0}, 
\pmatrix{0 & 0 & 1 \cr 1 & 0 & 0 \cr 0 & \omega^2 & 0}, 
\pmatrix{0 & 0 & \omega \cr \omega & 0 & 0 \cr 0 & 1 & 0},  
\pmatrix{0 & 0 & \omega \cr 1 & 0 & 0 \cr 0 & \omega & 0}, \nonumber \\
&:& \pmatrix{0 & 0 & 1 \cr \omega & 0 & 0 \cr 0 & \omega & 0}, 
\pmatrix{0 & 0 & \omega^2 \cr \omega^2 & 0 & 0 \cr 0 & \omega & 0}, 
\pmatrix{0 & 0 & \omega^2 \cr \omega & 0 & 0 \cr 0 & \omega^2 & 0}, 
\pmatrix{0 & 0 & \omega \cr \omega^2 & 0 & 0 \cr 0 & \omega^2 & 0}.
\end{eqnarray}

\newpage
\bibliographystyle{unsrt}

\end{document}